\documentclass[amsmath,twocolumn,showpacs,showkeys
]{revtex4}

\usepackage{amsmath,amsfonts,amsthm,amscd,amssymb}

\usepackage{epsf}
\usepackage{epsfig}

\def\a{\alpha}
\def\b{\beta}

\def\d{\delta}

\def\ep{\epsilon}

\def\@{\partial_}

\def\negenspace{\kern-1.1em}

\def\sqr#1#2{{\vcenter{\hrule height.#2pt\hbox{\vrule width.#2pt
height#1pt \kern#1pt \vrule width.#2pt}\hrule height.#2pt}}}
\def\square{\mathchoice\sqr64\sqr64\sqr{4.2}3\sqr{3.0}3}

\usepackage{graphicx, epic, eepic, color}

\definecolor{wichtig}{rgb}{1,0,0} 
\definecolor{folge}{rgb}{0,0,1} 
\definecolor{liste}{rgb}{0,0.7,0} 

\definecolor{dark-green}{rgb}{0,0.7,0}
\definecolor{dark-blue}{rgb}{0,0.2,0.5}
\definecolor{med-blue}{rgb}{0,0.7,1}
\definecolor{mblue}{rgb}{0,0.2,1}
\definecolor{cnc}{rgb}{0.8,0,0}
\definecolor{light-red}{rgb}{1,0.8,0.8}
\definecolor{dark-yellow}{rgb}{1,0.8,0}
\definecolor{light-blue}{rgb}{0.8,0.9,1}
\definecolor{verylight-blue}{rgb}{0.93,0.95,1}
\definecolor{light-yellow}{rgb}{1,0.9,0.8}
\definecolor{grey}{gray}{0.88}

\begin{document}
\title{Nonlocal Modification of Newtonian Gravity}

\author{Hans-Joachim Blome}
\email{blome@fh-aachen.de}
\affiliation{Department of Aerospace Technology,
Aachen University of Applied Sciences, Hohenstaufenallee 6,
52064 Aachen, Germany}
\author{Carmen Chicone}
\email{chiconec@missouri.edu}
\affiliation{Department of Mathematics, University of
Missouri, Columbia, Missouri 65211, USA}
\author{Friedrich W. Hehl}
\email{hehl@thp.uni-koeln.de}
\affiliation{Institute for Theoretical Physics, University of Cologne,
50923 K\"oln, Germany}
\affiliation{Department of Physics and Astronomy,
University of Missouri, Columbia, Missouri 65211, USA}
\author{Bahram Mashhoon}
\email{mashhoonb@missouri.edu}
\affiliation{Department of Physics and Astronomy,
University of Missouri, Columbia, Missouri 65211, USA}

\begin{abstract} 
  The Newtonian regime of a recent nonlocal extension of general
  relativity (GR) is investigated. Nonlocality is introduced via a
  scalar ``constitutive'' kernel in a special case of the
  translational gauge theory of gravitation, namely, the teleparallel
  equivalent of GR. In this theory, the nonlocal aspect of gravity
  simulates dark matter. A nonlocal and nonlinear generalization of
  Poisson's equation of Newtonian gravitation is presented. The
  implications of nonlocality for the gravitational physics in the
  solar system are briefly studied.
\end{abstract}



\date{10 March 2010, {\it file ModNewton26.tex}}

\maketitle

\section{Introduction}

The Poisson equation of Newtonian gravitation,
\begin{equation}\label{Poisson}
\nabla^2\Phi(t,\mathbf{x})=4\pi G \rho(t,\mathbf{x})
\end{equation}
is a consequence of the inverse-square force law, which is ultimately
based on solar-system observations that originally led to Kepler's
laws of planetary motion. Einstein's gravitational field equations
have generalized equation (\ref{Poisson}) into a consistent
relativistic framework that is in good agreement with present
solar-system data \cite{Shapiro,Will:2005,Everitt}. Nevertheless, on small
laboratory scales, for instance, questions remain regarding the
validity of the inverse-square law of gravitation and hence equation
(\ref{Poisson}); at present, efforts continue on resolving such
experimental problems
\cite{Adelberger:2003,Hoyle:2004,Adelberger:2006,Kapner:2006}. This
paper is about deviations from the inverse-square force law on
galactic scales in order to resolve the problem of the flat rotation
curves of spiral galaxies.

An essential component in the conceptual development of general
relativity (GR) is the way Lorentz invariance is employed to describe
what accelerated observers measure. Lorentz invariance is a
fundamental symmetry and refers to measurements of ideal inertial
observers that move uniformly forever on rectilinear timelike
worldlines; therefore, an assumption is required to relate these ideal
inertial observers to actual observers that are all non-inertial
(i.e., accelerated). The special theory of relativity uses the
postulate of locality, namely, the assumption that an accelerated
observer is {\it pointwise} inertial. The hypothesis of locality is
known to be an approximation \cite{Einstein,Weyl}; in fact, its domain
of applicability is limited to motions with sufficiently low
accelerations. The locality principle is also an essential ingredient
of Einstein's heuristic principle of equivalence that is the
cornerstone of general relativity. Nonlocal special relativity is a
generalization of the standard theory that goes beyond the locality
postulate and involves a certain average over the past worldline of
the observer \cite{MashhoonNonlocal}. The principle of equivalence of
inertial and gravitational masses implies a general connection between
inertia and gravitation; therefore, one would expect that the
nonlocality of accelerated observers in Minkowski spacetime would
entail a nonlocal theory of gravitation \cite{Bahram:2007}. However, a
direct nonlocal generalization of GR has not been possible; that is,
the highly local nature of Einstein's principle of equivalence
apparently prevents a straightforward nonlocal generalization of
GR. On the other hand, gauge theories of gravitation are in general
{\it less restrictive} \cite{vdH,Erice95}; hence, in principle, a
nonlocal generalization of GR can be constructed within the gauge
approach to gravitation. Indeed, in recent papers
\cite{nonlocal,NonLocal}, a nonlocal generalization of Einstein's
theory of gravitation has been presented on the basis of the
teleparallel equivalent of GR \cite{HehlHeld}. In the simplest
possibility, nonlocality is introduced via a scalar kernel. In this
approach to nonlocal gravity, nonlocality can persist in the Newtonian
limit of the theory.

To arrive at this limit in the linear approximation, it has been
assumed, in addition, that the scalar kernel ${\cal K}(x,y)$ is a
{\it universal} function of $ x-y$ and $x$ is supposed to be in the future
of $y$ to maintain causality \cite{nonlocal,NonLocal}. In this case,
the nonlocal gravitational field equations reduce to
\begin{equation}\label{Ba2}
  G_{\mu\nu}(x)+\int{\cal K}(x,y)G_{\mu\nu}(y)d^4y=\kappa T_{\mu\nu}\,,
\end{equation}
cf.\ Eq.~(62) of \cite{NonLocal}. Here $G_{\mu\nu}$ is the linear
Einstein tensor, $\kappa=8\pi G/c^4$ and $T_{\mu\nu}$ is the
energy-momentum tensor of matter ($\partial_\nu T^{\mu\nu}=0$). In
this gravitational background test particles and light rays
respectively follow timelike and null geodesics of the metric tensor
$g_{\mu\nu}=\eta_{\mu\nu}+h_{\mu\nu}$, where $\eta_{\mu\nu}$ is the
Minkowski metric tensor given by diag$(1, -1, -1, -1)$ and
$h_{\mu\nu}$ is the linear perturbation away from flat
spacetime. Greek indices run from 0 to 3, while Latin indices run from
1 to 3.
 
It is useful to move the nonlocal term to the right side of Eq.\
(\ref{Ba2}) via the Liouville-Neumann method of successive
substitutions. Let us introduce iterated kernels ${\cal K}_n$ given
by $ {\cal K}_1(x,y)= {\cal K}(x,y)$ and
\begin{equation}
  {\cal K}_{n+1}(x,y)
  =\int {\cal K}(x,z){\cal K}_n(z,y)d^4z\,.\label{Ba3}
\end{equation}
Inspection of Eq.\ (\ref{Ba3}) reveals that in each iterated kernel $
{\cal K}_{n}(x,y)$, with $n>1$, causality is preserved so that $x$ is
in the future of $y$, but in general ${\cal K}_{n}(x,y)$ is no longer
a function of $x-y$. This is therefore the case for the reciprocal
kernel ${\cal R}(x,y)$ as well,
\begin{equation}\label{Ba4}
-{\cal R}(x,y)=\sum_{n=1}^\infty {\cal K}_n (x,y)\,.
\end{equation}
Thus Eq.\ (\ref{Ba2}) can be written as
\begin{equation}\label{Ba5}
  G_{\mu\nu}(x)= \kappa\left[ T_{\mu\nu}(x) +\int{\cal R}(x,y)
    T_{\mu\nu}(y)d^4y\right]\,,
\end{equation}
so that the nonlocal theory in this approximation is equivalent to GR
but with an additional source term. In fact, the nonlocal aspect of
gravity can appear as {\it dark matter} given by the integral
transform of $T_{\mu\nu}$ by the {\it causal} reciprocal kernel $\cal
R$. In this paper we take the view that $\cal R$ must be determined
from observation; for instance, lensing observations of colliding
clusters of galaxies---such as in the case of the Bullet Cluster
\cite{Clowe:2006}---could provide clues regarding the nature of the
full time-dependent reciprocal kernel (cf.\ Section III).

In the Newtonian limit ($c\rightarrow \infty$), retardation effects
can be neglected and hence we can assume that each iterated kernel in
Eq.\ (\ref{Ba3}) is proportional to $\delta(x^0-y^0)$. It then follows
from Eq.\ (\ref{Ba3}) that
\begin{eqnarray}
{\cal R}(x,y)=\delta(x^0-y^0)q(\mathbf{x}-\mathbf{y})\,,
\end{eqnarray}
where $q$ is the spatial convolution kernel. Using this limiting form
of the kernel in Eq.\ (\ref{Ba5}), we find in the Newtonian limit the
nonlocal Poisson equation \cite{nonlocal,NonLocal}
\begin{equation}\label{Newton}
\nabla^2\Phi=4\pi G\left[\rho(t,\mathbf{x})+\rho_{\rm D}(t,\mathbf{x})\right]\,,
\end{equation}
where the ``density of dark matter'' $\rho_{\rm D}$ is given by
\begin{equation}\label{dark}
  \rho_{\rm D}(t,\mathbf{x})=\int q(\mathbf{x}-\mathbf{y})
  \rho(t,\mathbf{y})d^3y\,.
\end{equation}
Here $q$ is a universal function that is independent of the nature of
the source \cite{nonlocal,NonLocal}. This simplifying assumption is
relaxed in Section II, where we discuss the general form of the
nonlocal kernel in the Newtonian limit.

This paper is based on the assumption that there is no actual dark
matter. According to the approximation scheme employed in
\cite{nonlocal,NonLocal}, the nonlocal aspect of the gravitational
interaction acts like dark matter of density $\rho_{\rm D}$ that is
linearly related to the actual matter density via the kernel $q$ as in
equation (\ref{dark}). Consider, for example, the circular motion of
stars in the disk of a spiral galaxy in connection with the observed
flat rotation curves in such galaxies (see, for instance,
\cite{Carignan,Salucci} and references therein). At radius $r$
outside the bulge, the Newtonian acceleration of gravity for such a
star is nearly $v_0^2/r$, where $v_0$ is a constant speed. Poisson's
equation then implies that the corresponding density of ``dark''
matter must be $v_0^2/(4\pi G r^2)$. Extending this $\rho_{\rm D}$ to
a spherical distribution of ``dark'' matter by assumption, the result
can be compared with Eq.\ (\ref{dark}): neglecting the extended nature
of the galactic bulge and setting
$\rho(t,\mathbf{y})=M\delta(\mathbf{y})$, where $M$ is the effective
galactic mass, we find
\begin{equation}\label{qernel}
q(\mathbf{x}-\mathbf{y})=\frac{1}{4\pi\lambda}
\frac{1}{|\mathbf{x}-\mathbf{y}|^2}\,,
\end{equation}
where $\lambda=GM/v_0^2$ is of the order of 1\,kpc. The universality
of the nonlocal kernel implies that $\lambda$ must be a constant and
hence $M\propto v_0^2$. The resulting nonlocal modification of
Poisson's equation (\ref{Newton})--(\ref{qernel}) has been previously
discussed in connection with the Tohline-Kuhn scheme
\cite{Tohline,Kuhn,Jacob1988}. In particular, for a point source
$\rho(t,\mathbf{x})=M\delta(\mathbf{x})$, Eqs.\
(\ref{Newton})--(\ref{qernel}) imply that
\begin{equation}\label{pot1}
\Phi(t,\mathbf{x})=-\frac{GM}{|\mathbf{x}|}
+\frac{GM}{\lambda}\ln\left(\frac{|\mathbf{x}|}{\lambda}\right)\,.
\end{equation}
This coincides with Tohline's original suggestion regarding a
modification of Newton's law of gravitation in order to account for
the flat rotation curves of spiral galaxies \cite{Tohline}. A lucid
and enlightening account of the Tohline-Kuhn approach is contained in
the review paper of Bekenstein \cite{Jacob1988}.

It is clear from this brief account that, as shown in detail in
\cite{nonlocal,NonLocal}, the Tohline-Kuhn extension of Newtonian
gravitation to the realm of galaxies can be naturally embedded within
a nonlocal generalization of GR.  However, the Tohline-Kuhn scheme
disagrees with the empirical Tully-Fisher law \cite{TullyFisher}. The
Tully-Fisher relation involves a correlation between the luminosity of
a spiral galaxy and the corresponding asymptotic speed $v_0$. This
relation, combined with other empirical data regarding mass-to-light
ratio, roughly favors $M\propto v_0^4$, instead of $M\propto v_0^2$
that follows from the Tohline-Kuhn scheme. Various aspects of this
issue have been discussed in
\cite{Jacob1988,Jacob2004,Milgrom}. To go beyond the
Tohline-Kuhn scheme, we discuss a generalization of Eqs.\
(\ref{Newton})--(\ref{dark}) in Section II, where the Newtonian limit
of nonlocal gravity is discussed in detail. It is hoped that this more
general treatment of the Newtonian limit could help in the resolution
of the discrepancy with the Tully-Fisher law.

\section{Newtonian limit of nonlocal gravity}

The linear approximation in nonlocal gravity involves a linear
perturbation away from Minkowski spacetime. Consider a background
Minkowski spacetime with global inertial coordinates
$x^\a=(ct,\mathbf{x})$. The gravitational potentials are given by the
tetrad field $e_\mu{}^\nu(x)$ such that
\begin{equation}\label{Ba11}
  e_\mu{}^\a={\d}_\mu ^\a+\psi^\a{}_\mu\,,\quad e^\mu{}_\a
=\d^\mu _\a -\psi^\mu{}_\a\,,
\end{equation}
where $\psi_{\mu\nu}$ is proportional to $G/c^2$. The indices are
raised and lowered by means of the Minkowski metric tensor
$\eta_{\a\b}$. The gravitational field strength is then given by
\begin{equation}\label{Ba12}
  C_{\mu\nu\rho}=\psi_{\rho\nu,\mu}  -\psi_{\rho\mu,\nu}\,.
\end{equation}
We define the modified field strength $\frak{C}_{\mu\nu\rho}$ via
\begin{align}\label{Ba13}
\frak{C}_{\mu\nu\rho}=C_{\mu\nu\rho}+\psi_{[\mu\nu],\rho}
&+\eta_{\mu\rho}\left(\psi_{,\nu}-\psi_{\sigma\nu,}{}^\sigma\right)\nonumber\\
&-\eta_{\nu\rho}\left(\psi_{,\mu}-\psi_{\sigma\mu,}{}^\sigma\right)\,,
\end{align}
where $\psi=\eta_{\a\b}\psi^{\a\b}$. Thus both $C_{\mu\nu\rho}$ and
$\frak{C}_{\mu\nu\rho}$ are antisymmetric in their first two
indices. The coordinate components of the metric tensor are given by
\begin{equation}\label{Ba14}
g_{\mu\nu}=\eta_{\a\b}e_\mu{}^\a
e_\nu{}^\b=\eta_{\mu\nu}+\psi_{\mu\nu}
+\psi_{\nu\mu}\,.
\end{equation}
The nonlocal gravitational field equations in this linear approximation
scheme then reduce to \cite{nonlocal,NonLocal}
\begin{equation}\label{Ba15}
  G_{\mu\nu}(x)+\eta^{\rho\sigma}\int\frac{\partial K(x,y)}{\partial
    x^\rho}\frak{C}_{\mu\sigma\nu}(y)d^4y=\kappa T_{\nu\mu}(x)\,,
\end{equation}
where $G_{\mu\nu}$ is the linearized Einstein tensor in terms of
$g_{\mu\nu},T_{\mu\nu}$ is the energy-momentum tensor of mass-energy
in Minkowski spacetime and $\partial_\mu T^{\nu\mu}=0$. Equation
(\ref{Ba15}) corresponds to Eq.\ (60) of \cite{NonLocal}. If
$K(x,y)={\cal K}(x,y)$, then Eq.\ (\ref{Ba15}) reduces to Eq.\
(\ref{Ba2}), since
$G^{\mu\nu}=\partial_\sigma\frak{C}^{\mu\sigma\nu}$. In this case,
$T_{\mu\nu}$ must be symmetric and hence the antisymmetric part of
$\psi_{\mu\nu}$ does not participate in gravitational dynamics and can
thus be neglected. In general, however, the energy-momentum tensor is
{\it not} symmetric; therefore, Eq.\ (\ref{Ba15}) contains sixteen
field equations for the sixteen components of $\psi_{\mu\nu}$.

The kernel $K(x,y)$ could in general depend upon the structure of the
source in a manner that is consistent with the linear approximation
scheme.  This situation is considered in this section for the nonlocal
modification of Newtonian gravity. However, in previous work
\cite{nonlocal,NonLocal}, this possibility was neglected for the sake
of simplicity and it was assumed instead that $K(x,y)$ is some
universal function of $x-y$.  

To approach the Newtonian limit of the nonlocal theory, we tentatively
assume that the dynamics in Eq.\ (\ref{Ba15}) is dominated by the
Newtonian potential $\Phi$ such that
\begin{equation}\label{Ba16}
\psi_{00}=\psi_{11}=\psi_{22}=\psi_{33}=\frac{1}{c^2}\Phi
\end{equation}
and $\psi=-2\Phi/c^2$, while the other components of $\psi_{\mu\nu}$
may be neglected in the Newtonian limit. This assumption corresponds
to the circumstance that as in GR, one expects that the main effects
would be associated with a diagonal spacetime metric of the form
$g_{\mu\nu}=\eta_{\mu\nu}+h_{\mu\nu}$, where
$h_{\mu\nu}=2c^{-2}\Phi\,\text{diag}(1,1,1,1)$. In this case, we find
from Eq.\ (\ref{Ba13}) that
\begin{equation}\label{Ba17}
c^2\frak{C}_{0j0}=-2\partial_j\Phi\,;
\end{equation}
moreover, as in GR, $c^2G_{00}=2\nabla^2\Phi$. Hence, with
$T_{00}=\rho c^2$ and 
\begin{equation}\label{Ba18}
K(x,y)=\d(x^0-y^0)k(\mathbf{x},\mathbf{y})\,,
\end{equation}
where retardation effects have been neglected, the Newtonian limit of
Eq.\ (\ref{Ba15}) is of the form
\begin{equation}\label{Ba19}
  \nabla^2\Phi(\mathbf{x})+\sum_i\int\frac{\partial
    k(\mathbf{x},\mathbf{y})}{\partial x^i}\frac{\partial\Phi
    (\mathbf{y})}{\partial y^i}d^3y=4\pi G\rho(\mathbf{x})\,.
\end{equation}
This is a more general form of equations (\ref{Newton}) and
(\ref{dark}); furthermore, for simplicity we have suppressed any
temporal dependence. Equation (19) reduces to
Eqs. (\ref{Newton})--(\ref{dark}) if
$k(\mathbf{x},\mathbf{y})=k'(\mathbf{x}-\mathbf{y})$, for which the
reciprocal kernel is $q(\mathbf{x}-\mathbf{y})$; however, as pointed
out in \cite{nonlocal,NonLocal}, the kernel could in general depend
upon the Weitzenb\"ock invariants at $x$ and $y$. For the case under
consideration, these are
\begin{align}\label{Ba20}
  c^4\,C^{\mu\nu\rho}C_{\mu\nu\rho}&=6\eta^{\a\b}
  \frac{\partial\Phi}{\partial x^\a}\frac{\partial\Phi}{\partial x^\b}\,,\\
  c^4\,C^{\mu\nu\rho}C_{\rho\nu\mu}&=3\eta^{\a\b}\label{Ba21}
  \frac{\partial\Phi}{\partial x^\a}\frac{\partial\Phi}{\partial x^\b}\,, \\
  c^4\,C^{\mu\nu}{}_\nu C_{\mu\rho}{}^\rho&=9\left(\frac{\partial
      \Phi}{\partial x^0}\right)^2 -\left(\mathbf{\nabla}\label{Ba22}
    \Phi\right)^2\,.
\end{align}
It follows that in the limiting case ($c\rightarrow \infty$) under
consideration, the Weitzenb\"ock invariants all reduce to the square
of $|\mathbf{\nabla}\Phi|$, which is the magnitude of the Newtonian
gravitational acceleration. Hence, one may express the kernel as
\begin{equation}\label{Ba23}
  k(\mathbf{x},\mathbf{y})=k'(\mathbf{x}-\mathbf{y})+k''\left(\mathbf{x}
    -\mathbf{y};\frac{|\nabla_{\mathbf{y}}\Phi|}{|\nabla_{\mathbf{x}}
      \Phi|}\right)\,,
\end{equation}
so that $ k(\mathbf{x},\mathbf{y})$ depends on the structure of the
source, but is otherwise consistent with the linear approximation
scheme.

Equations (\ref{Ba19}) and (\ref{Ba23}) imply that
\begin{equation}\label{Ba24}
  \nabla_{\mathbf{x}}^2\Phi+\int k'(\mathbf{x},\mathbf{y})
\nabla_{\mathbf{y}}^2\Phi d^3y  =4\pi G(\rho+\rho_\Phi)\,,
\end{equation}
where $\rho_\Phi$ is defined by
\begin{equation}\label{Ba25}
\rho_\Phi(\mathbf{x})=-\frac{1}{4\pi G}
\sum_i\int \frac{\partial k''}{\partial x^i}\frac{\partial\Phi}
{\partial y^i}d^3y\,.\end{equation}
We recall that $q(\mathbf{x}-\mathbf{y})$ is reciprocal to
$k'(\mathbf{x}-\mathbf{y})$; therefore,
\begin{equation}\label{Ba26}
\nabla^2\Phi=4\pi G\{\rho+\rho_\Phi +\int q(\mathbf{x}-\mathbf{y})
[\rho(\mathbf{y})+\rho_\Phi(\mathbf{y})]d^3y \}\,.
\end{equation}
In the absence of $\rho_\Phi$, Eq.\ (\ref{Ba26}) is equivalent to
equations (\ref{Newton}) and (\ref{dark}). However, Eqs.\
(\ref{Ba24})--(\ref{Ba26}) contain a more general treatment of the
Newtonian limit of the nonlocal theory. Such a treatment is necessary
in order to help resolve observational problems associated with the
empirical Tully-Fisher relation \cite{TullyFisher}.  

Equation (\ref{Ba26}) is a {\it nonlinear} integro-differential
relation for the Newtonian potential $\Phi$. It is clear from Eq.\
(\ref{Ba23}) that scaling $\Phi$ by a constant factor leaves the
kernel invariant. Thus $\Phi$ given by Eq.\ (\ref{Ba26}), despite the
nonlinearity of this equation, will be linear in the gravitational
constant $G$, as would be expected on physical grounds. Moreover, as
in Newton's theory, the potential $\Phi$ can be determined from the
modified Poisson equation only up to an additive constant. Solutions
of equations (\ref{Ba19}) and (\ref{Ba23}), or equivalently
Eq. (\ref{Ba26}), are not known at present; therefore, in the
following sections we resort to the discussion of the solutions of the
{\it linear} part of the modified Poisson equation.

\section{Origin of kernel $q$}

We return to the study of equations (\ref{Newton}) and
(\ref{dark}). The main feature of these equations is the existence of
a {\it linear} relation between the potential $\Phi$ and matter
density $\rho$; that is,
\begin{equation}\label{pot2}
\Phi(t,\mathbf{x})=G\int\chi(\mathbf{x},\mathbf{y})\rho(t,\mathbf{y})d^3y\,.
\end{equation}
The Green function $\chi$ can in this case be simply obtained from
Eq.\ (\ref{pot1}), namely, $\chi$ is a function of $|\mathbf{x}-\mathbf{y}|$
and is given by
\begin{equation}\label{Green}
  \chi(\mathbf{x},\mathbf{y})=-\frac{1}{|\mathbf{x}-\mathbf{y}|}
  +\frac{1}{\lambda}\ln\left(\frac{|\mathbf{x}-\mathbf{y}|}
    {\lambda} \right)\,.
\end{equation}
It can be easily verified that $\chi$ is a solution of
\begin{equation}\label{solutionof}
\nabla^2_{\!\mathbf{x}}\chi(\mathbf{x},\mathbf{y})=4\pi
\left[\delta(\mathbf{x}-\mathbf{y})+q(\mathbf{x}-\mathbf{y}) \right]\,.
\end{equation}
One can develop {\it potential theory} (see, for instance,
\cite{Kellogg}) for nonlocal gravity on the basis of equations
(\ref{pot2})--(\ref{Green}). Moreover, the force of gravity per unit
test mass is given by
\begin{equation}\label{force}
  \hspace{-1pt}  
-\mathbf{\nabla}\Phi=-G\int\left[\frac{\mathbf{x}-\mathbf{y}}{|\mathbf{x}
      -\mathbf{y}|^3}
    +\frac{1}{\lambda} \frac{\mathbf{x}-\mathbf{y}}{|\mathbf{x}
      -\mathbf{y}|^2}\right]\rho(t,\mathbf{y})d^3y\,.
\end{equation}

The integral form of equation (\ref{Newton})---as well as its
generalization in equation (\ref{Ba26})---can be obtained using Green's
theorem; this is the subject of Appendix A.

In this paper, we take the tentative view that $q(\mathbf{r})$ must
ultimately be determined via observation. That is, this nonlocal
``Newtonian'' aspect of gravity, just as the local Newtonian
inverse-square force law, is a feature of the gravitational
interaction deducible from experience. Thus there is no fundamental
basis at present for the determination of the specific form of the
nonlocal kernel other than the concordance of equation (\ref{Newton})
with observational data. As pointed out in \cite{NonLocal}, the
convolution theorem for Fourier integrals may be employed to determine
$q$ using Eq.\ (\ref{dark}) once $\rho$ and $\rho_{\rm D}$ are {\it
  completely} known. However, this expectation is unrealistic at
present. Since Newton's time, various modifications of the
inverse-square force law have been contemplated \cite{North};
similarly, we can investigate how the potential (\ref{pot1}) would
change if the kernel (\ref{qernel}) is modified.

Let us first consider a kernel of the form
\begin{equation}\label{lkernel}
q(\mathbf{r})=\frac{1}{4\pi\lambda}\frac{1}{r^2+\ell_{0}^2}\,,
\end{equation}
where $r=|\mathbf{r}|$ and $\ell_{0}$ is a constant length parameter such
that for $\ell_{0}\ne 0$, Eq.\ (\ref{lkernel}) is, unlike Eq.\
(\ref{qernel}), singularity-free. Integrating the corresponding
Eq.\ (\ref{solutionof}), it is straightforward to show that
the analog of Eq.\ (\ref{pot1}) is in this case
\begin{equation}\label{lPhi}
\Phi=-\frac{GM}{r}+\frac{GM}{\lambda} \left[ \text{ln}\left( \frac{
({r^2+\ell_{0}^2})^{1/2}}{\lambda} \right)
+\frac{\ell_{0}}{r}\tan^{-1}\left(\frac{r}{\ell_{0}}\right) \right]\,,
\end{equation}
which reduces to Eq.\ (\ref{pot1}) for $\ell_{0}=0$. We note that the term
in square brackets goes to $1+\text{ln}(\ell_{0}/\lambda)$ for
$r\,\rightarrow\,0$; therefore, the logarithmic singularity in Eq.\
(\ref{pot1}) is avoided by the introduction of $\ell_{0}\neq 0$.  Next,
let
\begin{equation}\label{exponr}
q(\mathbf{r})=\frac{1}{4\pi\lambda}\frac{1}{r^2} e^{-r/L_{0}}\,,
\end{equation}
where $L_{0}$ is a constant length that renders the integral of
Eq.\ (\ref{exponr}) finite over all space. This is necessary to ensure that
the total mass of ``dark'' matter is finite; as pointed out in
\cite{NonLocal}, the total ``dark matter mass'' is infinite if
equation (\ref{qernel}) is taken to be valid for
$|\mathbf{x}-\mathbf{y}|\rightarrow\infty$. But empirical data are not
available beyond galaxy clusters and it is rather likely that
Eq.\ (\ref{qernel}) must be modified for sufficiently large
$|\mathbf{x}-\mathbf{y}|$. As before, it is possible to integrate
equation (\ref{solutionof}) in this case and the result is
\begin{eqnarray}\label{modpot}
\Phi&=&-\frac{GM}{r}+\frac{GM}{\lambda}\left[1+\frac{L_{0}}{r}\left(
e^{-r/L_{0}}-1\right)\right.\nonumber\\&&\left.
+\,\text{Ei}\left(-\frac{r}{L_{0}}\right)
-C-\ln\left(\frac{\lambda}{L_{0}} \right)\right]\,,
\end{eqnarray}
which reduces to Eq.\ (\ref{pot1}) for $L_{0}=\infty$. Here $C=0.577\dots$ is
the Euler constant and we use for the exponential integral function
$\text{Ei}\,(x)$ the expression
\begin{equation}\label{eiei}
\text{Ei}\,(-x)=C+\ln x+\sum_{n=1}^\infty \frac{(-1)^nx^n}{n\cdot
  n!}\,,\quad x>0
\end{equation}
and the asymptotic expansion
\begin{equation}\label{eiei2}
\text{Ei}\,(-x)=e^{-x}\sum_{n=1}^\infty
(-1)^n\frac{(n-1)!}{x^n}\,,\quad x\,\rightarrow\,\infty\,;
\end{equation}
see the first formula in (8.214) and formula (8.125) on page 927 of
Ref.\ \cite{R+G}. It is important to note that for
$r\,\rightarrow\,\infty$, $\Phi$ has a {\it constant} value in this
case.

It is clear from these considerations that variations in the simple
form of the kernel (\ref{qernel}) can lead to complicated expressions
for the gravitational potential. It is therefore interesting to
consider possible unique characterizations of this kernel.  In this
connection, we recall that in Newton's theory, the exterior
gravitational potential of a point mass is proportional to $1/r$ and
satisfies Laplace's equation. In fact, the fundamental harmonic
solution of Laplace's equation in $n$-dimensional Euclidean space is
$1/r^{n-2}$ for $n>2$ and $\ln r$ for $n=2$. In the $n=4$ case, this
result has a natural analog in Minkowski spacetime with inertial
coordinates $(ct,x,y,z)$, namely, $\Box W=0$, where
\begin{equation}\label{box}
W^{-1}=-c^2(t-t_0)^2+(x-x_0)^2+(y-y_0)^2+(z-z_0)^2\,.
\end{equation}
Here $\Box$ is the d'Alembertian operator defined by $\Box
=-\eta^{\a\b}\partial_\a\partial_\b$. For an interesting discussion of
such solutions and their singularities, see chapter IX of Synge
\cite{Synge}.

Let us note, for instance, that Eq.\ (\ref{qernel}) satisfies
\begin{equation}\label{wave=q2}
\nabla^2q=8\pi\lambda q^2\,.
\end{equation}
That is, up to a constant factor, $q(\mathbf{r})$ is a
time-independent solution of the semilinear wave equation
\cite{Derrick}
\begin{equation}\label{box1}
\square\,\varphi=\varphi^2\,.
\end{equation}
It is demonstrated in Appendix B that there is a one-parameter family
of nonzero spherically symmetric solutions of Eq.\ (\ref{wave=q2})
that vanishes together with all of their
derivatives as $r\rightarrow\infty$. These solutions, as discussed in
detail in Appendix B, behave as
\begin{equation}
  \frac{1}{4\pi\lambda}\left(\frac{1}{r^2}\pm\frac{C_0}{r^{2+\sigma}}
  \right)
\end{equation}
for $r\rightarrow\infty$ , where $\sigma=(\sqrt{17}-3)/2$ and $C_0$ is
an arbitrary constant parameter. With a suitable choice of $C_0$,
these latter solutions would also be consistent with galactic data.

Equations (\ref{wave=q2}) and (\ref{box1}) make it possible to
contemplate appropriate generalizations of Eq.\ (\ref{qernel}). For
instance, the invariance of Eq.\ (\ref{wave=q2}) under spatial
translations indicates that
\begin{equation}\label{1/r2}
\frac{1}{4\pi \lambda}\frac{1}{(x-x_0)^2+(y-y_0)^2+(z-z_0)^2}
\end{equation}
is also a solution of Eq.\ (\ref{wave=q2}) that reduces to
$q(\mathbf{r}), \mathbf{r}=(x,y,z)$, for
$\mathbf{r}_0=0$. Furthermore, it follows in a similar way from the
scalar field equation (\ref{box1}) that time-dependent kernels can be
constructed via Lorentz transformations. Consider, for instance, a
pure boost in the $x$ direction with speed $v$; then,
\begin{equation}\label{1/r2boost}
\frac{1}{4\pi \lambda}\frac{1}{\gamma^2(x+vt)^2+y^2+z^2}
\end{equation}
is a solution of Eq.\ (\ref{box1}) that reduces to $q(\mathbf{r})$ for
$v=0$. Here $\gamma$ is the Lorentz factor corresponding to speed
$v$. This means that one could construct reciprocal kernels involving
two events $(ct,\mathbf{r})$ and $(ct',\mathbf{r}')$ using functions of
the form
\begin{equation}\label{1/r2boost'}
  \frac{1}{4\pi \lambda}\frac{1}{\gamma^2[(x-x')+v(t-t')]^2+(y-y')^2
    +(z-z')^2}
\end{equation}
together with an appropriate causal ordering of the events.

In the rest of the paper, we simply employ kernel (\ref{qernel}).

\section{Extended Spherical Source}

According to the inverse-square force law, a homogeneous spherical
distribution of matter attracts an external particle as if the mass of
the sphere were concentrated at its center. However, this important
result of Newtonian gravitation would no longer hold in general with a
modified force law. To illustrate this point, equations (\ref{pot2})
and (\ref{Green}) can be used to evaluate the {\it exterior}
gravitational potential for any source distribution. Specifically, let
us consider a spherically symmetric mass distribution of radius $R_0$
such that
\begin{equation}\label{Mass}
M=4\pi\int_0^{R_0}\rho(r)r^2dr\,.
\end{equation}
At a spacetime position $(t,\mathbf{X})$ {\it exterior} to the {\it
  static} source, $R=|\mathbf{X}|>R_0$,
\begin{equation}\label{pot3}
\Phi(t,\mathbf{X})=-\frac{GM}{R}+\frac{2\pi
  G}{\lambda}\int_0^{R_0}F(R,r)
\rho(r)r^2dr\,.
\end{equation}
Here, the first term is due to the fact that in Newtonian gravitation
the exterior potential of any spherically symmetric distribution can
be replaced at its center by a point source, whose mass is equal to
the total mass of the spherical distribution. Moreover, in Eq.\
(\ref{pot3}), $F(R,r)$ is given by
\begin{equation}\label{FRr}
F(R,r)=\int_0^\pi \ln\left( \frac{
({R^2+r^2-2Rr\cos\theta})^{1/2}}{\lambda}\right)\sin\theta d\theta\,,
\end{equation}
since the symmetry of the configuration makes it possible to choose
the z axis to be along the position vector $\mathbf{X}$. It is
straightforward to show that
\begin{equation}\label{FRr1}
F(R,r)=2\ln\left(\frac{R}{\lambda} \right)+f(\frac{r}{R})\,,
\end{equation}
where for $\epsilon=r/R<1$,
\begin{equation}\label{fepsilon}
f(\ep)=-1+\frac{1}{2\ep}\left[(1+\ep)^2\ln(1+\ep)-(1-\ep)^2\ln(1-\ep) \right]\,.
\end{equation}
Thus Eq.\ (\ref{pot3}) can be written as
\begin{eqnarray}
\Phi(t,\mathbf{X})&=&-\frac{GM}{R}+\frac{GM}{\lambda}\ln\left(\frac{R}{\lambda}
\right)\nonumber \\&&+\frac{2\pi
G}{\lambda}\int_0^{R_0}f(\frac{r}{R})\rho(r)r^2dr\,, \label{pot4}
\end{eqnarray}
which is the sum of the contribution of a
point mass $M$ as in Eq.\ (\ref{pot1}) and an extra term due to the
extension of the source. One can show that
\begin{equation}\label{expansion}
f(\ep)=\frac 13 \ep^2 +\frac{1}{30}\ep^4+O(\ep^6)
\end{equation}
using the following relation that is valid for $|x|<1$,
\begin{equation}\label{ln}
\ln(1+x)=\sum_{n=1}^\infty(-1)^{n+1}\frac{x^n}{n}\,.
\end{equation}
To get an explicit result, let us assume, for the sake of simplicity,
that $\rho(r)=\rho_0$ is a constant. Then, the integral in Eq.\
(\ref{pot4}) can be evaluated analytically using the formulas (2.729)
on page 205 of Ref.\ \cite{R+G}. In any case, the dominant terms can
also be calculated directly from Eq.\ (\ref{expansion}) and the end
result is
\begin{eqnarray}
\Phi&=&-\frac{GM}{R}+\frac{GM}{\lambda}\ln\left(\frac{R}{\lambda}
\right)\nonumber \\ &&+\frac{GM}{10\lambda}\left(
\frac{R_0}{R}\right)^2\left[1+\frac{1}{14}\left(\frac{R_0}{R}
\right)^2+\cdots \right]\,.\label{pot5}
\end{eqnarray}

The force of gravity per unit test mass is conservative and is given
by $-\mathbf{\nabla}\Phi$; for Eq.\ (\ref{pot5}), this points in the
direction of the source and has a magnitude
\begin{equation}\label{force*}
\frac{d\Phi}{dR}=\frac{GM}{R^2}+\frac{GM}{\lambda R}\left[1-\frac 15
\left(\frac{R_0}{R} \right)^2 -\frac{1}{35}\left(\frac{R_0}{R}
\right)^4\cdots \right]\,.
\end{equation}
Here, the quantity in brackets is close to unity, since $R > R_0$;
therefore, we may conclude that the extended form of a nearly
homogeneous spherical source does not significantly alter the main
physical results of the Tohline-Kuhn scheme.

\section{Solar-system effects}

It is interesting to search for evidence of nonlocal gravity in the
solar system. As a first step in this endeavor, let us consider
gravitational physics in the solar system using a Tohline-Kuhn
gravitational potential of the form
\begin{equation}\label{Ba54}
\Phi=-\frac{GM}{r}+\frac{GM}{\lambda}\ln\frac{r}{\lambda'}\,,
\end{equation}
where the gravitational source is at the origin of coordinates. Here
$\lambda'$ is assumed to be a galactic-scale length. For
$\lambda=10\;\text{kpc},\,2\;\text{A.U.}/\lambda\approx10^{-9}$, hence the
logarithmic term in Eq.\ (\ref{Ba54}) is expected to be a very small
perturbation of the Newtonian potential. The following preliminary
considerations are based on the fact that in the nonlocal
generalization of GR under consideration here, light rays and test
particles move along null and timelike geodesics, respectively.

\subsection{Time Delay}

In Newtonian gravity the potential vanishes at infinity by convention.
However, the logarithmic term in Eq. (54) is assumed to vanish at the
radial distance $\lambda'$. While this is of no consequence in
Newtonian gravity, it matters here as the spacetime interval depends
on $\Phi$. Consider, for instance, the gravitational time delay
$\Delta$ between events $P_1:(ct_1,\mathbf{r}_1)$ and
$P_2:(ct_2,\mathbf{r}_2)$ when a light signal travels from $P_1$ to
$P_2$. Let $L=|\mathbf{r}_2-\mathbf{r}_1|$ and $\ell:0\rightarrow L$
be the distance along a straight line from $P_1$ to $P_2$; then,
$\Delta=t_2-(t_1+L/c)$ is given by \cite{Ciufolini}
\begin{equation}\label{Ba55}
\Delta=-\frac{2}{c^3}\int_{P_1}^{P_2}\Phi d\ell\,.
\end{equation}
It follows from a detailed calculation that
\begin{align}
  \Delta={}&\frac{2GM}{c^3}\ln
  \frac{r_2+\hat{\mathbf{n}}\cdot\mathbf{r}_2}
  {r_1+\hat{\mathbf{n}}\cdot\mathbf{r}_1}
 \nonumber\\
  & -\frac{2GM}{c^3\lambda}\left\{(\hat{\mathbf{n}}\cdot\mathbf{r}_2)
    \ln\frac{r_2}{\lambda'}-(\hat{\mathbf{n}}\cdot\mathbf{r}_1)
\ln\frac{r_1}{\lambda'}-L\right.\nonumber\\&\left.
 +A\left[\tan^{-1}\left(\frac{\hat{\mathbf{n}}\cdot{\mathbf{r}}_2}{A}
      \right)-\tan^{-1}\left(\frac{\hat{\mathbf{n}}\cdot{\mathbf{r}}_1}{A}
      \right)\right]\right\}\,.\label{Ba56}
\end{align}
Here\vspace{-8pt}
\begin{equation}\label{Ba57}
\hat{\mathbf{n}}=\frac 1L (\mathbf{r}_2-\mathbf{r}_1)\,,\qquad
A=\frac 1L |\mathbf{r}_1\times\mathbf{r}_2|\,.
\end{equation}
The net result is a sum of the Shapiro time delay together with the
contribution of the logarithmic term in the potential. The relative
magnitude of these delays is expected to be similar as in Eq.\
(\ref{Ba54}).

In the rest of this section, only the derivative of the potential is
involved; therefore, $\lambda'$ drops out of our calculations. For
instance, in the gravitational shift of the frequency of light only
the difference in the potential $\Phi$ at two spatially separated
events is significant.

\subsection{Deflection of Light}

The net deflection angle $D$ of a light ray due to a point mass $M$
with potential $\Phi$ is given by twice the Newtonian expectation in
the first post-Newtonian approximation, which for Eq.\ (\ref{Ba54})
works out to be
\begin{equation}\label{Ba58}
D=\frac{4GM}{c^2\zeta}+\frac{2\pi GM}{c^2\lambda}\,,
\end{equation}
where $\zeta$ is the distance of the closest approach. The bending
angle is thus slightly larger than the Einstein angle by a
constant. However, the extra deflection is not expected to remain
constant for an extended source (cf.\ Section IV).

The effect of the logarithmic potential is $\sim \!10^{-12}$ of the
Einstein angle for the bending of light by the Sun.

\subsection{Perihelion Precession}

The gravitational force due to potential (\ref{Ba54}) is radial and
conservative.  Therefore, the perturbing influence of the logarithmic
term in Eq.\ (\ref{Ba54}) on Keplerian orbits is such that the orbit
remains planar and the orbital angular momentum is unchanged. Let
$(r,\phi)$ be polar coordinates in the orbital plane and consider an
unperturbed Keplerian ellipse given by
\begin{equation}\label{Ba59}
r=\frac{a(1-e^2)}{1+e\cos \hat{\phi}}\,,
\end{equation}
where $a$ is the semimajor axis of the ellipse, $e$ is its
eccentricity, $\hat{\phi}=\phi-g$ and $g$ is the argument of the
pericenter. Under the influence of the radial perturbing acceleration
$-(GM/\lambda)r^{-1}$, the orbital elements of the {\it osculating}
ellipse vary in accordance with the Lagrange planetary equations
\cite{Danby}. In this case, we find
\begin{align}\label{Ba60}
\frac{da}{dt}=&-\frac{2\omega a^2 e}{\lambda(1-e^2)^{3/2}}
(1+e\cos\hat{\phi})\sin\hat{\phi}\,,\\
\frac{dg}{dt}=&\frac{\omega a}{\lambda e(1-e^2)^{1/2}}
(1+e\cos\hat{\phi})\cos\hat{\phi}\,,\label{Ba61}
\end{align}
where $\omega$ is the Keplerian frequency of the osculating ellipse
($\omega^2=GM/a^3$). Moreover, $[GMa(1-e^2)]^{1/2}$ is the magnitude
of the specific orbital angular momentum and remains constant.
Let us note here that only positive square roots are considered in
this paper. 

The elements of the osculating ellipse change slowly
according to Eqs.\ (\ref{Ba60}) and (\ref{Ba61}); therefore, it is
natural to average the right-hand sides of these equations over the
fast orbital motion with period $T=2\pi/\omega$. That is, we define
the average of a quantity $Q$ to be
\begin{equation}\label{Ba62}
<Q>\;=\frac{1}{T}\int_0^T Qdt\,,
\end{equation}
so that
\begin{equation}\label{Ba63}
<Q>\;=\frac{(1-e^2)^{3/2}}{2\pi}\int_0^{2\pi}
\frac{Q\,d\phi}{(1+e\cos\hat{\phi})^2}\,,
\end{equation}
since $r^2d\phi/dt=[GMa(1-e^2)]^{1/2}$ for the unperturbed orbit. It
follows that $<da/dt>\;=0$, so that the semimajor axis remains unchanged on the
average. This is also the case for the orbital eccentricity due to the
constancy of the angular momentum. Thus the ellipse keeps its shape on
the average but precesses, since
\begin{equation}\label{Ba64}
<\frac{dg}{dt}>\;=-\frac{\omega a}{2\lambda}P(e)\,,
\end{equation}
where
\begin{equation}\label{Ba65}
  P(e)=\frac{2}{e^2}\left[(1-e^2)^{1/2}-(1-e^2) \right]
\end{equation}
decreases from unity at $e=0$ to zero at $e=1$. Here we have
used the fact that
\begin{equation}\label{Ba66}
  \int_{\a_0}^{\a_0+2\pi}\frac{d\a}{1+e\cos\a}=\frac{2\pi}{(1-e^2)^{1/2}}\,.
\end{equation}
One can obtain the same result for the pericenter precession frequency from
the study of the variation of the Runge-Lenz vector \cite{LL}.

For the solar system, the resulting perihelion precession
\cite{Schmidt} is retrograde and for $\lambda=10\text{ kpc}$, it is
about $10^{-3}$ of Einstein's value for Mercury and about $2\times
10^{-2}$ for Earth, as there is more ``dark matter'' to influence the
outer orbits. The general relativistic contribution to the perihelion
precession of Mercury is known at present at the level of about one
part in a thousand; therefore, the possible contribution of the
logarithmic potential is hidden within the present measurement
error. Future improvements in such measurements may make it possible
to detect the influence of nonlocal gravity in the solar
system. Nevertheless nonlocal effects appear at present to be too
small to be detectable. For instance, the contribution of the
logarithmic term to the Pioneer anomaly is $\sim 10^{-4}$ of the
anomalous acceleration of the Pioneer spacecraft.

It appears that other anomalies in the solar system---such as the
flyby anomaly, the possible secular increase in the Astronomical Unit
and the increase in the eccentricity of Moon's orbit (see
Ref.\ \cite{A+Nieto})---are not directly affected by the conservative
perturbing force under consideration here. We should also mention that
solar-system deviations from GR can in principle be used to place
lower bounds on the constant lengthscales that appear in the
logarithmic term in Eq.\ (\ref{Ba54}).

\section{Discussion}

Starting from first principles, arguments have been advanced for a
nonlocal generalization of Einstein's theory of gravitation
\cite{MashhoonNonlocal,Bahram:2007,nonlocal,NonLocal}.  In such a
theory, the gravitational field is local, but satisfies nonlocal
integro-differential field equations. These are obtained from the
local field equations via a nonlocal ``constitutive'' ansatz, as
described in Appendix C within the general context of gauge theories
of gravitation that are less restrictive than GR and thus make it
possible to implement this procedure.

The Newtonian limit of the simplest nonlocal GR theory involving a
scalar constitutive kernel \cite{nonlocal,NonLocal} is studied in this
paper. It is shown that the theory reduces to a {\it nonlinear} and
nonlocal modification of Poisson's equation of Newtonian gravity. The
exploration of the nonlinear aspects of this equation is beyond the
scope of the present work; therefore, we ignore the nonlinear part of
this equation and concentrate on the simpler case of the linear
Poisson equation. This turns out to be equivalent to the Tohline-Kuhn
scheme of modified Newtonian gravity as an alternative to dark matter
\cite{Tohline,Kuhn,Jacob1988,Jacob2004,Milgrom}. Indeed, on galactic
scales, the nonlocal deviation of the gravitational interaction from
the inverse-square force law could be responsible for observational
data that have been attributed to the presence of dark matter. As a
preliminary step, we study some of the implications of the linear
nonlocal theory for observations within the solar system.

\appendix

\section{Integral form of Poisson's equation}

Consider a source density $\rho(\mathbf{x})$ with compact support in a
spatial volume $V$ bounded by the surface $S$. Using Eq.\ (\ref{Newton})
and
\begin{equation}\label{B1}
\nabla^2\frac{1}{|\mathbf{x}-\mathbf{x'}|}=-4\pi\delta(\mathbf{x}-\mathbf{x'})
\end{equation}
in Green's theorem, we find
\begin{equation}\label{B2}
\Phi(\mathbf{x})=-G{\cal I}+{\cal S}\,,\quad \mathbf{x}\in V\,,
\end{equation}
while
\begin{equation}\label{B3}
G{\cal I}={\cal S}\,,\quad\mathbf{x}\notin V\,.
\end{equation}
Here
\begin{equation}\label{B4} {\cal I}=\int_V\frac{\rho(\mathbf{x'})+\rho_{\rm
      D}(\mathbf{x'})}{|\mathbf{x}-\mathbf{x'}|}d^3x'
\end{equation}
and $\cal S$ is the surface integral
\begin{equation}\label{B5}
{\cal S}=\frac{1}{4\pi}\oint_S\left[\frac{1}{{\cal R}}\frac{\partial
    \Phi}{\partial n'}
  -\Phi(\mathbf{x'}) \frac{\partial}{\partial n'}\left(\frac{1}{{\cal
        R}} 
\right) \right]dS\,,
\end{equation}
where ${\cal R}=|\mathbf{x}-\mathbf{x'}|$, $\partial\Phi/\partial n'
:=(\nabla_{\mathbf{x'}}\Phi)\cdot \widehat{\mathbf{n}}'$ and
$\widehat{\mathbf{n}}'$ is the unit vector normal to the boundary
surface $S$.

In Eq.\ (\ref{B3}), where $\mathbf{x}\notin V$, and assuming that $q$ and
$\rho$ are continuous functions, the order of integration in $\cal
I$---when Eq.\ (\ref{dark}) is taken into account---may be interchanged
such that
\begin{equation}\label{B6}
{\cal I}=\int_V K(\mathbf{x},\mathbf{x'})\rho(\mathbf{x'})d^3x'\,,
\end{equation}
where $K$ is given by
\begin{equation}\label{B7}
  K(\mathbf{x},\mathbf{x'})=\frac{1}{|\mathbf{x}-\mathbf{x'}|}+\int_V 
\frac{q(\mathbf{y}-\mathbf{x'} ) d^3y}{|\mathbf{x}-\mathbf{y}|}\,.
\end{equation}

If $\rho+\rho_{\rm D}$ is bounded for small $r=|\mathbf{x}|$, falls
off as $r^{-(2+\a)}$ with $\a>0$ for large $r$ and
$\Phi\,\rightarrow\,0$ as $r\,\rightarrow\,\infty$, then
\begin{equation}\label{B8}
  \Phi(\mathbf{x})=-G\int\frac{\rho(\mathbf{x'})+\rho_{\rm D}
(\mathbf{x'})}{|\mathbf{x}-\mathbf{x'}|}d^3x'\,. 
\end{equation}
To satisfy the conditions for Eq.\ (\ref{B8}), we note that for a smooth
source density $\rho(\mathbf{x})$ with compact support, $\rho_{\rm D}$
must be finite and $q(\mathbf{x})$ must fall off as $r^{-(2+b)}$ with
$b>0$ as $r\,\rightarrow\, \infty$, since it is simple to check that
\begin{equation}\label{B9}
\nabla^2\left(\frac{1}{r^b} \right)=\frac{b(b-1)}{r^{b+2}}\,.
\end{equation}

\section{Nonuniqueness of $q$}

We consider the time independent, spherically symmetric solutions of
the nonlinear equation
\begin{equation} \square \,u=2 u^2\end{equation} in spacetime. Using
spherical coordinates, the spherically symmetric, time-independent
solutions satisfy the ordinary differential equation
\begin{equation}\label{eq1}
u_{rr}+\frac{2}{r} u_r=2 u^2.
\end{equation}
In particular, we wish to determine the solutions that are
$C^\infty$-flat as $r\to \infty$.

The change of variables (see~\cite{B,G-H})
\begin{equation}
z=r^2 u, \qquad \tau = \ln r
\end{equation}
or, equivalently,
\begin{equation} z(\tau)=e^{2\tau} u(e^{\tau}), \qquad
  u(r)=\frac{1}{r^2} z(\ln r),\end{equation} transforms
equation~\eqref{eq1} to the autonomous ordinary differential equation
\begin{equation}\label{eq2} z''-3 z' +2 z =2 z^2,\end{equation}
where the prime signifies differentiation with respect to $\tau$.
\begin{figure}
\centerline{\includegraphics[width=20pc]{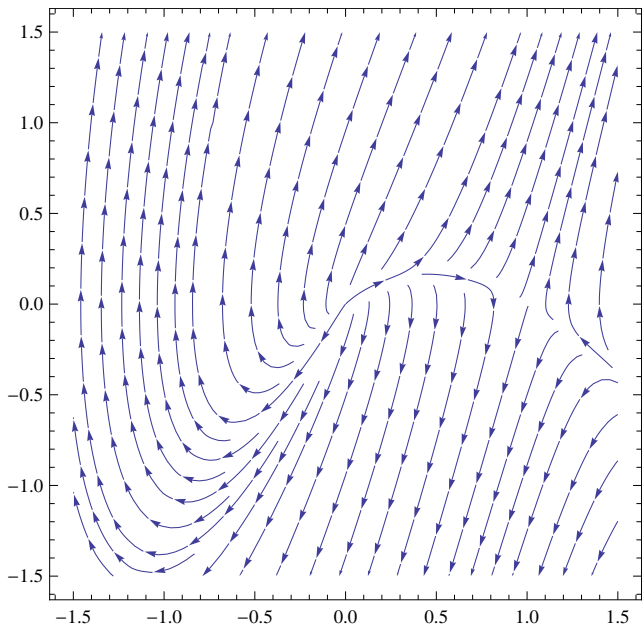}}
  \caption{Phase portrait of system~\eqref{eq3}, where $w$ is drawn
    versus $z$.\label{fig:pp}}
\end{figure}

In the phase plane (see Figure 1), the corresponding system
\begin{equation}\label{eq3} z'=w,\qquad w'=3 w-2 z+2 z^2 \end{equation}
has two rest points $(z,w)=(0,0)$ and $(z,w)=(1,0)$, which correspond
to the solutions
\begin{equation}u(r)=0, \qquad u(r)=1/r^2
\end{equation} 
of equation~\eqref{eq1}; these solutions might also be obtained by
inspection of this equation. Both of these solutions and all of their
derivatives vanish in the limit as $r\to \infty$; that is, both are
$C^\infty$-flat as $r\to \infty$.

By linearization at the point $(0,0)$ in the phase plane for
system~\eqref{eq3}, it follows that this point is a hyperbolic source
with corresponding spectrum $\{1,2\}$. The rest point $(1,0)$ is a
hyperbolic saddle with spectrum $\{\frac{1}{2}(3-\sqrt{17}),
\frac{1}{2}(3+\sqrt{17})\}$. Moreover, the stable manifold of this
rest point is tangent to the line with parametrization
\begin{align}\label{eq4}
s\mapsto \left(\begin{array}{c}
1\\
0
\end{array}
\right)+s \left(\begin{array}{c}
-1\\
\frac{1}{2}(\sqrt{17}-3)
\end{array}
\right)\,.
\end{align}

A solution of the system whose orbit lies on this stable manifold is
asymptotic to a corresponding solution of the linearization on the
linearized stable manifold (which is exactly the line in
display~\eqref{eq4}). That is, the asymptotic behavior of a solution
on the stable manifold is
\begin{align}\label{eq5}
\tau \mapsto \left(\begin{array}{c}
1\\
0
\end{array}
\right)\pm  e^{-\sigma (\tau+\tau_0)} \left(\begin{array}{c}
-1\\
\frac{1}{2}(\sqrt{17}-3)
\end{array}
\right),
\end{align}
where $ \sigma=\frac{1}{2}(\sqrt{17}-3) $ and $\tau_0$ is an arbitrary
real number.  In particular, the asymptotic behavior of the first
component of a solution on the stable manifold will be
\begin{equation}
z(\tau)\sim 1\pm e^{-\sigma(\tau-\tau_0)},
\end{equation}
which corresponds to
\begin{equation}
u(r)\sim \frac{1}{r^2}\pm \frac{e^{\sigma\tau_0}}{r^{2+\sigma}}.
\end{equation}
Because $\sigma>0$, the function $u$  decreases like $1/r^2$ as $r\to \infty$.

\section{Nonlocal Poincar\'e gauge theory}

In the traditional approach to general relativity (GR), the invariance
of the spacetime manifold under coordinate transformations is
emphasized. Thus curvilinear coordinate systems play a dominant role
in GR and observers that occupy fixed positions in space in a given
coordinate system are accelerated in general. The frame field
constructed from linearly independent vectors that are tangent to the
curvilinear coordinate lines are called {\it natural} or {\it
  coordinate} frames. Such a frame, which is not necessarily
orthonormal, is integrable by definition and hence {\it
  holonomic}. The gravitational field equations are {\it second-order}
local partial differential equations (PDEs) for the spacetime metric
in a given system of coordinates.

In the gauge approach to gravity, however, one considers arbitrary
frame fields that are in general non-integrable, that is, {\it
  anholonomic}. Taking advantage of the freedom afforded by the use of
anholonomic frames and the associated geometric concepts (such as
torsion), the gravitational field equations take the form of {\it
  first-order} local PDEs \cite{Schouten,vdH,Hartley,Erice95,JimNormal}.

It turns out that one can extend the {\it first-order local} field
equations to {\it nonlocal} ones via the introduction of a
``constitutive'' kernel as in the phenomenological electrodynamics of
media. In this way, a nonlocal generalization of Einstein's theory of
gravitation becomes possible by starting with the teleparallel
equivalent of GR rather than with GR itself \cite{NonLocal,nonlocal}.


In order to be able to construct a nonlocal generalization of
Einstein's gravitational theory, the gauge theory of translations was
recently employed in \cite{NonLocal,nonlocal}. Now, the gauge theory
of translations itself is a somewhat degenerate subcase of the gauge
theory of the Poincar\'e group, the so-called ``Poincar\'e gauge
theory of gravity''. In turn, the question arises whether the
nonlocal generalization of the translational gauge theory can be
extended to the Poincar\'e gauge theory.  This is, in fact,
the case. We will follow the method used in \cite{NonLocal,nonlocal}
for the translational gauge theory in the more general case of the
Poincar\'e gauge theory. We will use the notation and conventions of
\cite{NonLocal}.

Let the gauge Lagrangian of the underlying Riemann-Cartan spacetime
depend on coframe $e_i{}^\a$, torsion $T_{ij}{}^\a$ and curvature
$R_{ij}{}^{\a\b}=-R_{ij}{}^{\b\a}$; that is, ${\cal
  L}_{\text{grav}}={\cal L}_{\text{grav}}(e_i{}^\a,
T_{ij}{}^\a,R_{ij}{}^{\a\b})$. The matter Lagrangian ${\cal
  L}_{\text{mat}}$, with the matter field(s) $\Psi$, is supposed to be
minimally coupled to the geometry. Then the total Lagrangian reads
\begin{equation}\label{totLagr}
\hspace{-1pt}
{\cal L}_{\text{tot}}={\cal
  L}_{\text{grav}}(e_i{}^\a,T_{ij}{}^\a,R_{ij}{}^{\a\b})
+{\cal L}_{\text{mat}}(e_i{}^\a,\psi,D_i\psi)\,,
\end{equation}
with the independent field variables $e_i{}^\a$ (coframe),
$\Gamma_i{}^{\a\b}=-\Gamma_i{}^{\b\a}$ (Lorentz connection) and
$\Psi$ (matter field(s)). With the help of the two excitations 
\begin{equation}\label{excit}
{\cal H}^{ij}{}_\a=-2\frac{\partial{\cal L}_{\text{grav}} }{\partial T_{ij}{}^\a}
\quad\text{and}\quad {\cal H}^{ij}{}_{\a\b}=-2\frac{\partial
{\cal L}_{\text{grav}} }{\partial R_{ij}{}^{\a\b}}\,,
\end{equation}
the two field equations---the results of the variation of ${\cal
  L}_{\text{tot}}$ with respect to $e_i{}^\a$ and
$\Gamma_i{}^{\a\b}$---can be written as
\begin{align}\label{first}
D_j{\cal H}^{ij}{}_\a-{\cal E}_\a{}^i&=\Sigma_\a{}^i\,,\\
\label{second}
D_j{\cal H}^{ij}{}_{\a\b}-e^j{}_{[\a}{\cal H}^i{}_{|j|\b]}&=\tau_{\a\b}{}^i\,,
\end{align}
where $\Sigma_\a{}^i=\delta{\cal L}_{\text{mat}}/\delta e_i{}^\a$
denotes the canonical energy-momentum tensor density of the matter
field and $\tau_{\a\b}{}^i=\delta{\cal
  L}_{\text{mat}}/\delta\Gamma_i{}^{\a\b}=-\tau_{\b\a}{}^i$ denotes
the corresponding canonical spin (angular momentum) tensor density
(note that these definitions differ slightly from the ones in
Ref.\ \cite{HehlHeld}). The energy-momentum tensor density
$\Sigma_\a{}^i$ should not be confused with the torsion tensor
$T_{ij}{}^\a$.

In Eq.\ (\ref{first}), the energy-momentum tensor of the gauge fields
can be expressed as
\begin{equation}\label{gravEnergy}
{\cal E}_\a{}^i:=e^i{}_\a {\cal L}_{\text{grav}}-{\cal H}^{jk}{}_\a
T_{jk}{}^i -{\cal H}^{jk}{}_{\a\b}R_{jk}{}^{i\b}\,.
\end{equation}
On the other hand, for Eq.\ (\ref{second}) the spin of the gauge
fields is given by $e^j{}_{[\a}{\cal H}^i{}_{|j|\b]}$, which depends
only on the translational excitation; thus, it is very simple and we
have already substituted it directly into Eq.\ (\ref{second}).

This represents the general framework for the Poincar\'e gauge
theory. We still have to specify the explicit form of the gauge
Lagrangian. Following the general scheme of a Yang-Mills theory, we
assume that the Lagrangian is local and quadratic in torsion and
curvature. We denote the three irreducible pieces of the torsion by
$^{(I)}T_{ij}{}^\a$, for $I=1,2,3$, and the six irreducible pieces of
the curvature by $^{(K)}R_{ij}{}^{\a\b}$, for $K=1,2,...,6$; details can
be found in \cite{HehlHeld,Erice95}. Then the Lagrangian reads
\begin{align}\label{quadrLagr}\nonumber
  \stackrel{\hspace{-10pt}\text{loc}}{{\cal
      L}_{\text{grav}}}&=\frac{1}{2\kappa} \sqrt{-g}\Big[-e^i{}_\a
    e^j{}_\b \,^{(6)}R_{ij}{}^{\a\b} +\Lambda
    +T^{ij}{}_\a\\ & \hspace{-7pt}\times \sum_{I=1}^3
    b_I\,^{(I)}T_{ij}{}^\a\Big]
  +\frac{\sqrt{-g}}{2\xi}R^{ij}{}_{\a\b}\sum_{K=1}^6 c_K\,
  ^{(K)}R_{ij}{}^{\a\b},
\end{align}
where $\kappa$ is Einstein's gravitational constant, $\Lambda$ is the
cosmological constant and $\xi$ is the dimensionless coupling constant of
``strong gravity'', which is mediated via the propagating Lorentz
connection.  The constants $b_I$ and $c_K$ are dimensionless and
should be of order unity.

We compute the excitations from Eq.\ (\ref{quadrLagr}) by partial
differentiation according to Eq.\ (\ref{excit}):
\begin{align}\label{excitTr}
{\cal H}^{ij}{}_\a&=\frac{\sqrt{-g}}{\kappa} \sum_{I=1}^3 b_I
\,^{(I)}T^{ij}{}_\a\,,\\ \nonumber {\cal
  H}^{ij}{}_{\a\b}&=\frac{\sqrt{-g}}{\kappa}e^i{}_{[\a} e^j{}_{\b]}
+\frac{\sqrt{-g}}{\xi}\sum_{K=1}^6 c_K\, ^{(K)}R^{ij}{}_{\a\b}
\\ \label{excitRot}&= \stackrel{\text{lin}}{{\cal H}}{}^{ij}{}_{\a\b}\,+
\stackrel{\text{qu}}{{\cal H}}{}^{ij}{}_{\a\b}\,.
\end{align}
This is the quadratic {\it local} Poincar\'e gauge theory.

For later purposes, it is convenient to express the Lagrangian in
terms of the excitations: 
\begin{align}
{\cal L}_{\text{grav}}= &-\frac 12 \stackrel{\text{lin}}{{\cal
  H}}{}^{ij}{}_{\a\b}\,^{(6)}R_{ij}{}^{\a\b}
+\frac{\sqrt{-g}}{2\kappa}
\Lambda\nonumber\\ \label{Ham}
&-\frac 14 {\cal H}^{ij}{}_\a T_{ij}{}^\a
-\frac 14  \stackrel{\text{qu}}{{\cal H}}{}^{ij}{}_{\a\b}
R_{ij}{}^{\a\b}\,.
\end{align}
This Lagrangian will also be valid in the nonlocal case. We now
generalize the local ``constitutive relations'' (\ref{excitTr}) and
(\ref{excitRot}) to nonlocal ones, again as in
\cite{nonlocal,NonLocal}, by using an unknown scalar kernel
$\chi({x},{x}')$ and the world function $\Omega$ and its
derivatives for transporting tensors from ${x}'$ to ${x}$:
\begin{align}\label{excitNonl1}
{\cal H}^{ij}{}_k({x})&=-\frac{\sqrt{-g(x)}}{\kappa}
\int U( {x}, {x}')\Omega^{ii'} \Omega^{jj'}
\Omega_{kk'}\nonumber\\
&\hspace{-25pt}\times \chi( {x}, {x}')\sum_{I=1}^3 b_I
\,^{(I)}T_{i'j'}{}^{k'} (
{x}')\sqrt{-g(x')}d^4x',\\ \label{excitNonl2}
\stackrel{\text{lin}}{{\cal
    H}}{}^{ij}{}_{kl}( {x})&=\frac{\sqrt{-g(x)}}{\kappa}\,
\d^i_{[k}\d^j_{l]},\\
\label{excitNonl3}
\stackrel{\text{qu}}{{\cal
    H}}{}^{ij}{}_{kl}( {x})&=\frac{\sqrt{-g(x)}}{\xi} 
\int U( {x}, {x}')\Omega^{ii'} \Omega^{jj'}
\Omega_{kk'}\Omega_{ll'}\nonumber\\
&\hspace{-30pt}\times \chi( {x}, {x}')\sum_{K=1}^6 c_K
\, ^{(K)}R_{i'j'}{}^{k'l'}( {x}')\sqrt{-g(x')} d^4x',\\ 
\label{excitNonl4} {\cal
  H}^{ij}{}_{kl}&=\stackrel{\text{lin}}{{\cal H}}{}^{ij}{}_{kl}+
 \stackrel{\text{qu}}{{\cal H}}{}^{ij}{}_{kl}.
\end{align}

The final field equations will not be written down explicitly. We find
them as follows: We first substitute Eqs.\
(\ref{excitNonl1})--(\ref{excitNonl4}) into (\ref{Ham}) and
(\ref{gravEnergy}) and then into the field equations (\ref{first}) and
(\ref{second}); the last step involves the substitution of the new
Eq.\ (\ref{gravEnergy}), after Eq.\ (\ref{Ham}) is inserted, into Eq.\
(\ref{first}). In this way, we have a set of 16 + 24
integro-differential equations in terms of the variables
$e_i{}^\a,\Gamma_i{}^{\a\b}$ and $\Psi$.

\end{document}